\RequirePackage{fix-cm}
\documentclass[twocolumn,epjc3]{svjour3}  
\smartqed  
\RequirePackage{epsfig}
\RequirePackage{url}
\RequirePackage{hyperref}

\RequirePackage[normalem]{ulem} 
\usepackage{graphicx}
\usepackage{mathtools}
\RequirePackage{latexsym}
\RequirePackage{epsfig}
\RequirePackage{amsmath}
\RequirePackage{amssymb}
\RequirePackage{wasysym}
\RequirePackage{graphicx}
\RequirePackage{verbatim}
\RequirePackage{enumerate,mdwlist}
\RequirePackage[titletoc]{appendix}
\RequirePackage{amsfonts}
\RequirePackage{tikz} 
\usetikzlibrary{calc}
\RequirePackage{pgfplots}
\RequirePackage[export]{adjustbox}

\newcommand{\HCd}{\mathcal{H}}

\def\HCdt0{\tilde{\HCd}_{0}}

\newcommand\br{\begin{eqnarray}}
\newcommand\er{\end{eqnarray}}
\newcommand\be{\begin{equation}}
\newcommand\ee{\end{equation}}

\newcommand\bc{\begin{center}}
\newcommand\ec{\end{center}}

\newcommand\m{\mu}

\renewcommand\r{\rho}

\renewcommand\S{\Sigma}

\newcommand\twomat[4]{\left(\begD{array}{cc}  
{#1} & {#2} \\ {#3} & {#4} \end{array} \right)}

\RequirePackage{mathptmx}      
\RequirePackage{latexsym}
\RequirePackage[numbers,sort&compress]{natbib}
\newcommand{\Mpl}{M_{\textrm{Pl}}}

\def\S{\mathcal{S}}

\def\br{\bar\rho}

\def\doi{http://doi.org}

\def\r{\mathrm{r}}

\def\m{\mathrm{m}}

\journalname{}
\begin{document}
\title{Quintessential Inflation from Lorentzian Slow Roll} 


\author{D. Benisty\thanksref{e1,addr1,addr2}
        \and
E. I. Guendelman\thanksref{e2,addr1,addr2,addr3} 
}

\thankstext{e1}{e-mail: benidav@post.bgu.ac.il}
\thankstext{e2}{e-mail: guendel@bgu.ac.il}

\institute{Physics Department, Ben-Gurion University of the Negev, Beer-Sheva 84105, Israel \label{addr1}
\and
Frankfurt Institute for Advanced Studies (FIAS), Ruth-Moufang-Strasse~1, 60438 Frankfurt am Main, Germany \label{addr2}
\and
Bahamas Advanced Study Institute and Conferences, 4A Ocean Heights, Hill View Circle, Stella Maris, Long Island, The Bahamas  \label{addr3}}

\date{}

\maketitle

\begin{abstract}
From the assumption that the slow roll parameter $\epsilon$ has a Lorentzian form as a function of the e-folds number $N$, a successful model of a quintessential inflation is obtained, as succinctly studied in \cite{Benisty:2020xqm}. The form corresponds to the vacuum energy both in the inflationary and in the dark energy epochs and satisfies the condition to climb from small values of $\epsilon$ to $1$ at the end of the inflationary epoch. We find the corresponding scalar Quintessential Inflationary potential with two flat regions. Moreover, a reheating mechanism is suggested with numerical estimation for the homogeneous evolution of the universe. The suggested mechanism is consistent with the BBN bound.
\end{abstract}

\section{Introduction}
The inflationary paradigm is considered as a necessary part of the standard model of 
cosmology, since it provides the solution to the fundamental puzzles of the old Big 
Bang theory, such as the horizon, the flatness, and the monopole problems
\cite{Guth:1980zm,Guth:1982ec,Starobinsky:1979ty,Kazanas:1980tx,Starobinsky:1980te,Linde:1981mu,Albrecht:1982wi,Barrow:1983rx,Blau:1986cw}. It can be achieved through various mechanisms, for instance through the introduction of a scalar inflaton field \cite{Barrow:2016qkh,Barrow:2016wiy,Olive:1989nu,Linde:1993cn,Liddle:1994dx,Germani:2010gm,Kobayashi:2010cm,Feng:2010ya,Burrage:2010cu,Kobayashi:2011nu,Ohashi:2012wf,Cai:2014uka,Kamali:2016frd,Benisty:2017lmt,Middleton:2019bio,Dalianis:2018frf,Dalianis:2019asr,Qiu:2020qsq}. Almost twenty years after the observational evidence of cosmic acceleration the cause of this phenomenon, labeled as dark energy", remains an open question which challenges the foundations of theoretical physics: why there is a large disagreement between the vacuum expectation value of the energy momentum tensor which comes from quantum field theory and the observable value of dark energy density \cite{Weinberg:1988cp,Copeland:2006wr,Lombriser:2019jia,Merritt:2017xeh}. One way to parametrize dynamical dark energy is by using a scalar field or a "quintessence" model  \cite{Ratra:1987rm,Caldwell:1997ii,Benisty:2018qed}. In such a way that the cosmological constant gets replaced by a dark energy fluid with a nearly constant density today \cite{Zlatev:1998tr,Caldwell:1999ew,Chiba:1999ka,Bento:2002ps,Tsujikawa:2013fta}. For the slow roll approximation the scalar field behaves as an effective dark energy. The form of the potential is clearly unknown and many different potentials have been studied and confronted to observations.

These two regimes of accelerated expansion are treated independently. However, it is both tempting and economical to think that there is a unique cause responsible for a quintessential inflation
\cite{wetterich,murzakulov-etal,Copeland:2000hn,BuenoSanchez:2006fhh,Bento:2008yx,Guendelman:2015mva,Guendelman:2016kwj,Guendelman:2017mkj,Hossain:2014xha,Hossain:2014ova,Hossain:2014coa,Hossain:2014zma,Geng:2015fla,Staicova:2016pfd,Geng:2017mic,Kaganovich:2000fc,Akrami:2017cir,Dimopoulos:2017zvq,Benisty:2018gzx,Kiselev:2018iru,Staicova:2018ggf,Benisty:2018fja,Staicova:2018bdy,Benisty:2019tno,Benisty:2019bmi,Benisty:2019jqz,Benisty:2019pxb,Staicova:2019ksr,Lima:2019yyv,Benisty:2020vvm,Staicova:2020zwo} which refers to unification of both concepts using a single scalar field. Consistency of the scenario demands that the new degree of freedom, namely the scalar field, should not interfere with the thermal history of the Universe, and thereby it should be ``invisible'' for the entire evolution and reappear only around the present epoch giving rise to late-time cosmic acceleration.

While the standard approach is to introduce some potential with slow roll behavior, a successful model of quintessential inflation can be obtained from the ansatz of the slow roll parameter on as a function of the scale parameter itself (or the number of $e$-folds) \cite{Artymowski:2019jlh,Benisty:2019vej,Odintsov:2020thl}, especially, with a Lorentzian ansatz \cite{Benisty:2020xqm}. Here we calculate the complete evolution of the universe from this ansatz.

The plan of the work is as follows: In section \ref{sec:LorAnz} we introduce the ansatz for the slow roll parameter and the corresponding inflationary observables. Section \ref{sec:scal} finds the scalar potential that reads the same behavior. In section \ref{sec:reh} we introduce an example for reheating mechanism. Section \ref{sec:late} calculates the expansion of the universe with matter fields. Finally section \ref{sec:con} summarizes the results.

\section{Lorentzian ansatz}
\label{sec:LorAnz}
\begin{figure}
 	\centering
\includegraphics[width=0.45\textwidth]{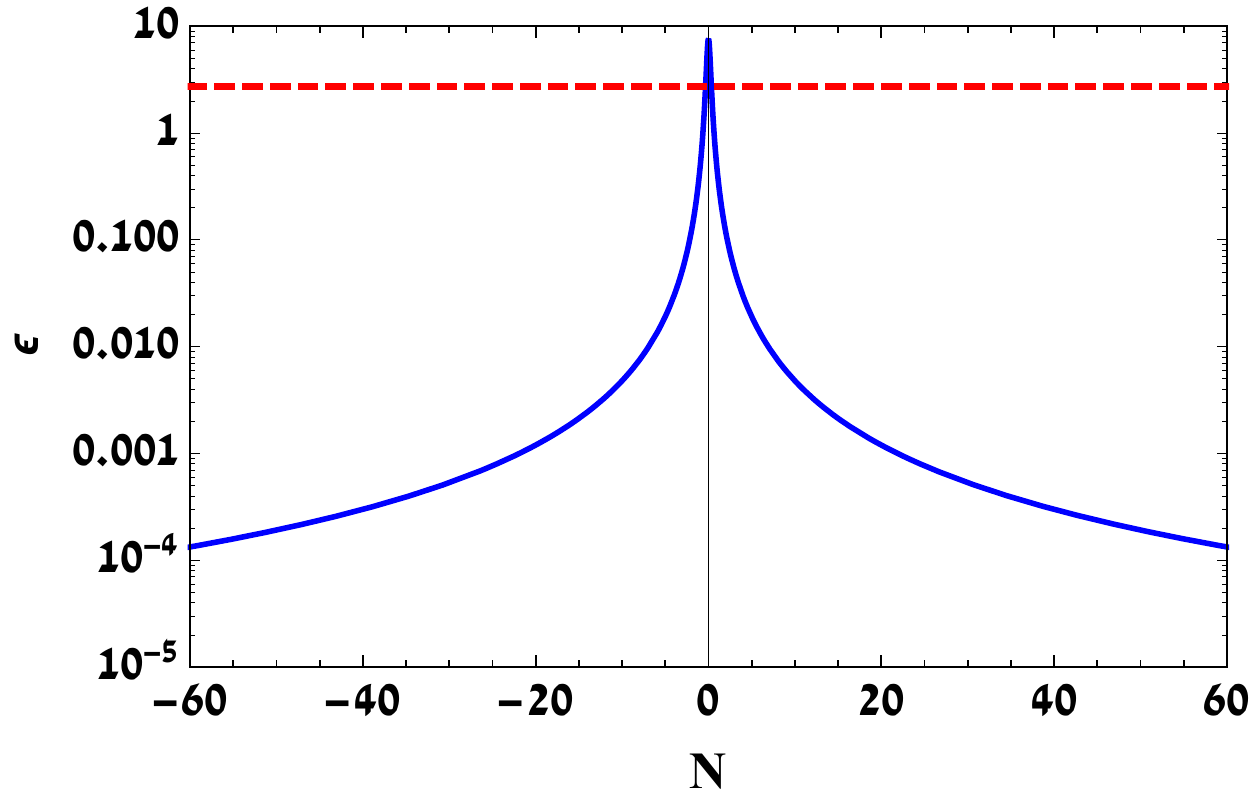}
\includegraphics[width=0.45\textwidth]{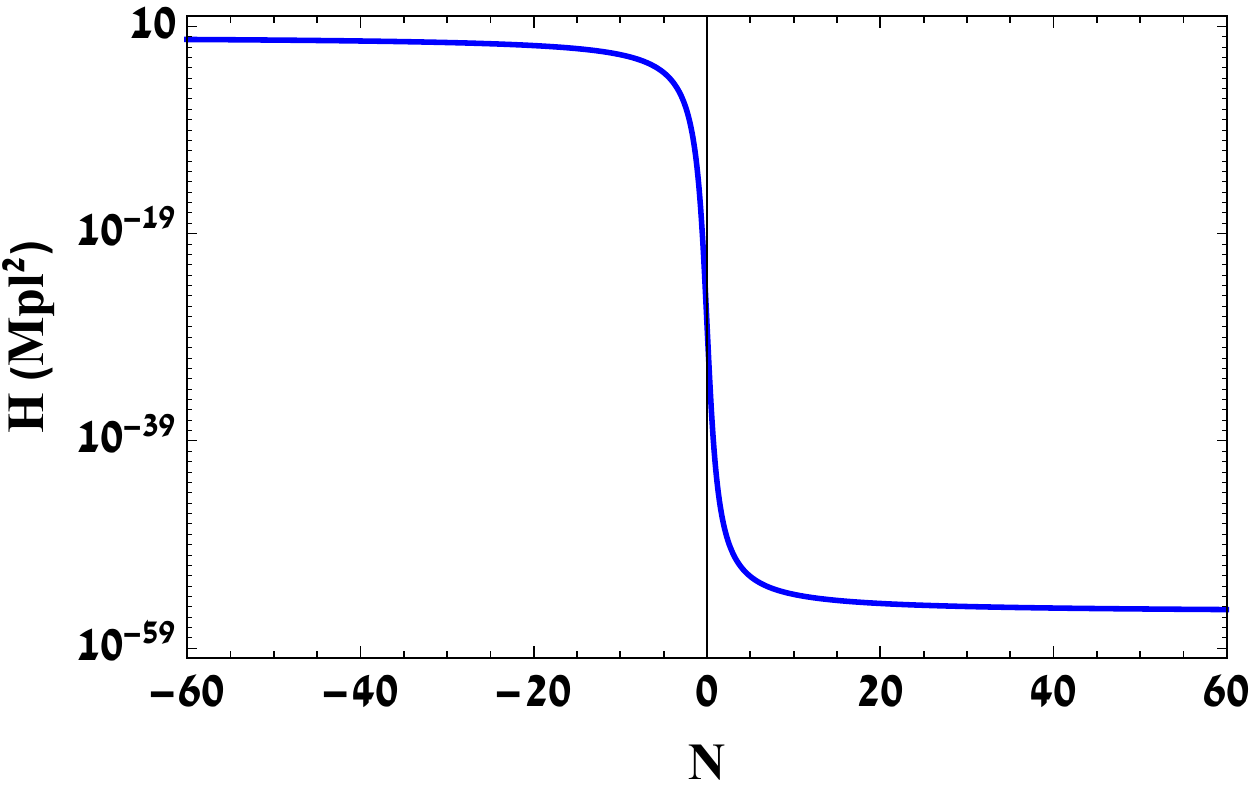}
\caption{{\it{The upper panel shows the slow roll parameter $\epsilon$ vs. the number of $e$-folds for the ansatz (\ref{eq:ans}), in a logarithmic scale. The lower panel shows the corresponding Hubble function of the vacuum vs. the number of $e$-folds.}}}
\label{fig1}
\end{figure}

In order to formulate an ansatz for the Hubble function that treats symmetrically both the early and late times we use the Lorentzian function for the slow roll parameter:
\begin{equation}
\epsilon(N) = \frac{\xi}{\pi}\frac{\Gamma/2 \, }{ N^2 + (\Gamma/2)^2}
\label{eq:ans}
\end{equation}
as a function of the number of $e$-folds $N = \log(a/a_i)$, where $a_i$ is the scale parameter at some time (which we may choose as the initial state of the inflationary phase). $\xi$ is the amplitude of the Lorentzian, $\Gamma$ is the width of the Lorentzian. In that way the $\epsilon$ parameter increases from the initial value to $1$ at the end of inflation,then continues to increase, peak and then decreases until it gets down to the value $1$ and this represents the beginning of a the new Dark Energy phase that will eventually dominate the late evolution of the Universe. The upper panel of Fig \ref{fig1} presents the qualitative shape of this behavior.

The dominant energy condition yields another bound on the coefficients. The equation of states $w$ is in the range $ |w| \leq 1$. From the relation $\epsilon = \frac{3}{2}\left(w + 1\right)$ we obtain the bound $ 0 \leq \epsilon \leq 3$. The ansatz for the vacuum energy evolution (\ref{eq:ans}) positive always, hence the lower bound is preserved. The largest value of the ansatz (\ref{eq:ans}) is $2 \xi /\pi  \Gamma $. From the the upper bound of $\epsilon$ we obtain the condition:
$\Gamma < 2 \xi/ 3\pi$. 

In general, the calculation of the above observables demands a detailed perturbation analysis. Nevertheless, one can obtain approximate expressions by imposing the slow-roll assumptions, under which all inflationary information is encoded in the slow-roll parameters. In particular, one first introduces \cite{Gong:2001he,Martin:2002vn,Leach:2002ar,Habib:2002yi,Casadio:2006wb,Martin:2013tda}:
\begin{equation}\label{slowRoll}
\epsilon_{n+1} = \frac{d}{dN} \log 
|\epsilon_n|,
\end{equation}
where $\epsilon_0\equiv H_{i}/H$ and $n$ is a positive integer. The slow roll parameters read:
\begin{equation*}
\epsilon \equiv \epsilon_1 = -\frac{H'}{H},\quad \epsilon_2 = \frac{H''}{H'}-\frac{H'}{H},
\end{equation*}
and so on. From the first slow roll parameter definition with the ansatz (\ref{eq:ans}), we obtain the solution:
\begin{equation}
H = \sqrt{\frac{\Lambda_0}{3}} \exp[-\frac{\xi}{\pi}\tan ^{-1}\left(\frac{2 N}{\Gamma }\right)]. 
\label{eq:hub}
\end{equation}
where $\Lambda_0$ is an integration constant. The Hubble function interpolates from the inflationary values $H_{-\infty}$ to the dark energy value $H_{+\infty}$ that corresponds to:
\begin{equation}
H_{\pm } = \sqrt{\frac{\Lambda_0}{3}}\exp^{\mp \xi/2}.
\end{equation}
The magnitude of the vacuum energy at the inflationary phase reads $10^{-8} Mpl^4$, while the magnitude of the vacuum energy at the present slowly accelerated phase of the universe is $10^{-120} Mpl^4$. From the Friedmann equations the values of the energy density is $3H^2$ in the Planck scale. Therefore, the coefficients of the model are:
\begin{equation}
\xi \approx 129,\quad \Lambda_0 = 1.7 \cdot 10^{-32} Mpl^4.
\end{equation} 
We calculate the other slow roll parameters using (\ref{slowRoll}):
\begin{equation}
\epsilon_2 = -\frac{8 N}{\Gamma ^2+4 N^2}, \quad
\epsilon_3 = \frac{1}{N}-\frac{8 N}{\Gamma ^2+4 N^2}.
\end{equation}
For $\Gamma \rightarrow 0$ all of the slow roll parameters with $n \geq 3$ yields the value $-1/N$. However in the general case, all of the slow parameters have small values if the $\epsilon_2$ is small.

As usual inflation ends at a scale factor $a_f$
where $\epsilon_1(a_f)=1$ and the slow-roll approximation breaks down. Therefore the end of inflation takes place when the number of $e$-folds read:
\begin{equation}
N_{f} =  \pm \sqrt{\frac{\Gamma}{4 \pi } (2 \xi-\pi  \Gamma)}
\end{equation}
The negative value of $N_{f}$ is the final state of the inflationary phase, while the positive value of $N_f$ is the initial value of the slow rolling Dark Energy at the late universe. Therefore, in order to calculate the inflationary observables, we must take the minus sing of $N_{f}$. we take Consequently the initial $N_i$ satisfies the condition: $N_f - N_i = \mathcal{N} \approx 50-60$,
where we impose $60\,e$-folds for the inflationary phase. Hence, the initial state of the inflationary phase reads:
\begin{equation}
N_i = - \sqrt{\frac{\Gamma}{4 \pi } (2 \xi-\pi  \Gamma)} - \mathcal{N}
\end{equation}
  \begin{figure*}
 	\centering
 	\includegraphics[width=0.95\textwidth]{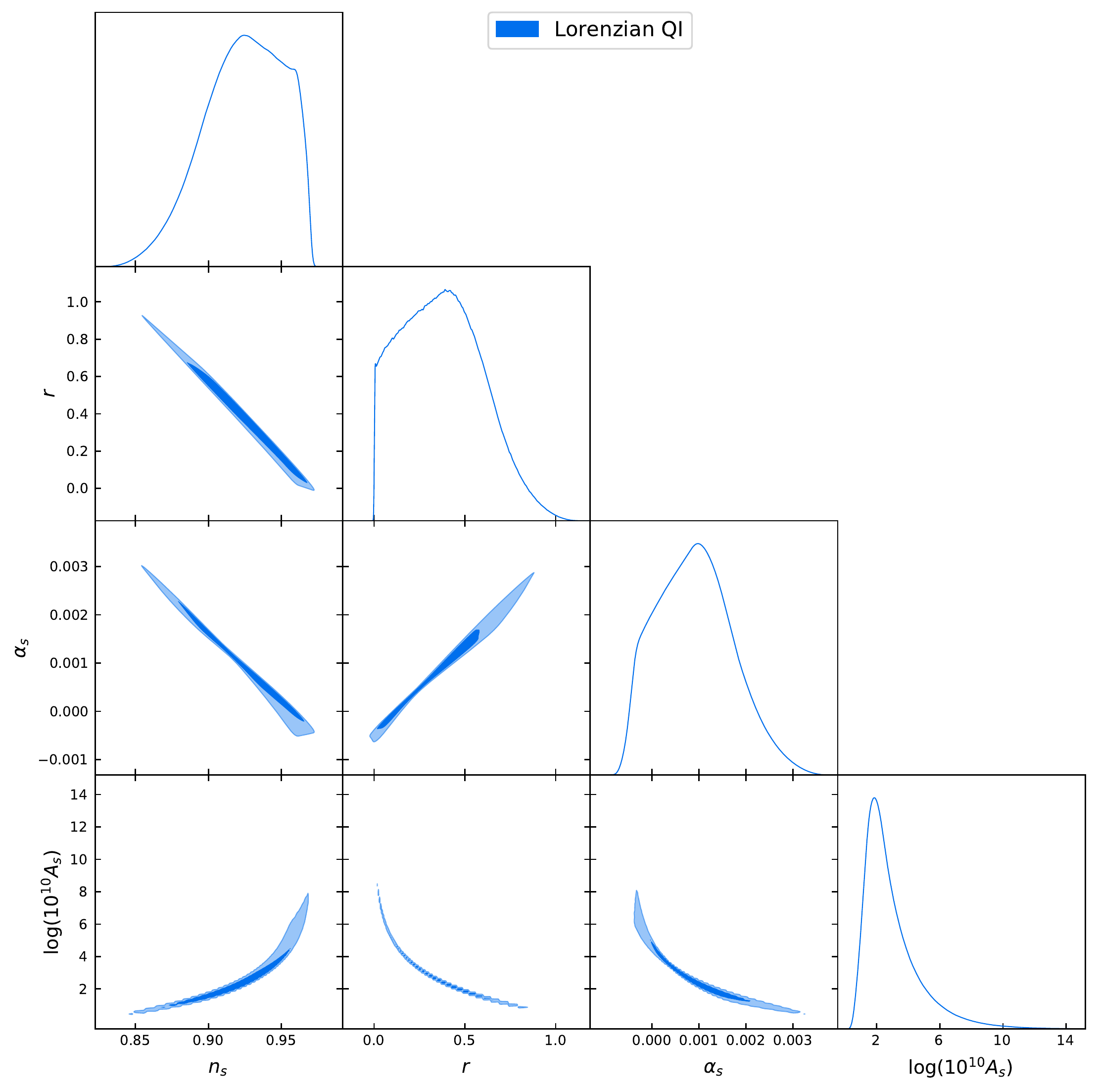}
\caption{{\it{The predicted scalar to tensor ratio $r$ and the running $\alpha_s$ vs. the primordial tilt $n_s$ of the model. The observation are well in the Planck 2018 bounds. }}}
\label{fig2}
\end{figure*}
The inflationary observables are expressed as \cite{Gong:2001he,Martin:2002vn,Leach:2002ar,Habib:2002yi,Casadio:2006wb,Martin:2013tda}:
\begin{equation}
r \approx 16\epsilon_1,
\end{equation}
\begin{equation}
n_\mathrm{s} \approx 1-2\epsilon_1-\epsilon_2,
\end{equation}
\begin{equation}
\alpha_\mathrm{s} \approx -2 \epsilon_1\epsilon_2-\epsilon_2\epsilon_3,
\end{equation}
\begin{equation}
\begin{split}
n_\mathrm{T} \approx -2\epsilon_1,    
\end{split}
\end{equation}
where all quantities are calculated at $N_{i}$. Therefore the tensor to scalar ratio and the primordial tilt give:
\begin{subequations}
\begin{equation}
r = \frac{32 \Gamma  \xi }{\pi  \Gamma ^2+4 \pi  N_i^2}, 
\end{equation}
\begin{equation}
n_s = \frac{\pi  \left(\Gamma ^2+4 N_i (N_i+2)\right)-4 \Gamma  \xi }{\pi  \left(\Gamma ^2+4 N_i^2\right)},
\end{equation}
\begin{equation}
\alpha_s = \frac{8 \left(\pi  \Gamma ^2-4 \pi  N_i^2+4 \Gamma  \xi  N_i\right)}{\pi  \left(\Gamma ^2+4 N_i^2\right)^2}
\end{equation}
\end{subequations}
 The power spectrum reads:
\begin{equation}
\begin{split}
A_s \approx \frac{H^2}{8\pi^2 \epsilon}.
\end{split}
\end{equation}
For $60\, e$-folds and $\Gamma = 0.1$ the observables read:
\begin{equation}
r = 0.0076, \quad n_s = 0.961754.
\end{equation}
These values in agreement with the latest $2018$ Planck data \cite{Aghanim:2018eyx,Akrami:2018odb}:
\begin{equation}
0.95 < n_s < 0.97, \quad r < 0.064
\end{equation}
Fig \ref{fig2} shows the predicted distribution of the observables \cite{getDist}. We assume a uniform prior: $N \in [50;70]$, $\xi \in [100;200]$, $\Gamma \in [0;1]$, with $10^7$ samples. We find the posterior yields:
\begin{equation}
r = 0.045^{+0.065}_{-0.053}, 
\end{equation}
\begin{equation}
n_s = 0.9624^{+0.0087}_{-0.011},
\end{equation}
\begin{equation}
\alpha_s = -\left(33^{+27}_{-30}\,\right)\cdot 10^{-5}, 
\end{equation}
\begin{equation}
\log\left(10^{10} A_s\right) = 3.49 \pm 2.98 \, 
\end{equation}
 The predicted distribution are in a good agreement with the recent Planck values.
\section{Scalar field dynamics}
\label{sec:scal}
The above ansatz is of general applicability in any inflation realization, whether this is driven by a scalar field, or it arises effectively from modified gravity, or from any other mechanism. In order to provide a more transparent picture let us consider a realization of these ideas in the context of a canonical scalar field theory $\phi$ moving in a potential $V\left(\phi \right)$. We consider the action
\begin{equation}
\mathcal{S} = \int d^4x
\sqrt{-g}\left[\frac{\Mpl^2}{2}R-\frac{1}{2}\partial^\mu\phi\partial_\mu \phi-V(\phi) \right]
 \label{action0}
\end{equation}
with $\Mpl$ the Planck mass, $\phi$ the scalar field, and $V(\phi)$ its potential. In the above action we have additionally considered the matter and radiation sectors $\mathcal{S}_{m}$ and $\mathcal{S}_{\gamma}$ respectively. These sectors can be neglected at the inflationary stage, however they will gradually play an important role, giving rise to the standard thermal history of the Universe
 and finally to the late-time accelerating phase. As usual, we focus on the case of a flat Friedmann-Robertson-Walker (FRW) geometry, with the metric:
\begin{equation}
\label{metric0}
 ds^2 = -dt^2 +a(t)^2\delta_{ij} dx^idx^j ~,
\end{equation}
where $a(t)$ is the scale factor. The Friedman equations are given by:
\begin{equation}\label{FriedInf}
 H^2  = \frac{8\pi G}{3}\left[\frac{1}{2} \dot{\phi}^2 + V(\phi)\right], \quad
 \dot{H} = -4\pi G\dot{\phi}^2,
\end{equation}
while the variation for the scalar field is
\begin{equation}
\label{KleinG}
\ddot{\phi} + 3 H \dot{\phi} + V'(\phi) = 0.
\end{equation}
Let us apply the ansatz in order to reconstruct a physical scalar-field potential that can generate the desirable inflationary observables. From the Friedmann equation (\ref{FriedInf}) that holds in every scalar-field inflation, we extract the 
following solutions:
\begin{equation}\label{conTran1}
\phi = \int_{0}^{N} \sqrt{-2 \frac{H'}{H}} \, dN, \quad
V(\phi) = H H'+3 H^2.
\end{equation}
with $8\pi G = 1$. From the integration of the Hubble parameter we get:
\begin{equation}
N = \frac{\Gamma}{2}   \sinh \left(\sqrt{\frac{\pi}{\xi \Gamma}}\phi\right),      
\label{VphiN}
\end{equation}
\begin{equation}
V(N) = \Lambda_0 e^{-\frac{2 \xi}{\pi }\tan^{-1}\left(\frac{2 N}{\Gamma }\right)} \left(1-\frac{2 \Gamma  \xi }{3 \pi  \Gamma ^2+12 \pi  N^2} \right).
\end{equation}
Expression (\ref{VphiN}) cannot be inversed, in order to find $N(\phi)$ and then through 
insertion into (\ref{VphiN}) to extract $V(\phi)$
analytically:  
\begin{equation}
\begin{split}
V(\phi) = \Lambda_0 e^{-\frac{2 \xi}{\pi} \tan ^{-1}\left(\sinh x \right)} \left(1-\frac{2 \xi }{3 \pi  \Gamma } \text{sech}^2 x \right). 
\end{split}
\label{FinPot}
\end{equation}
with $x \equiv \sqrt{\pi/\Gamma \xi} \phi$.

\begin{figure}[t!]
 	\centering
\includegraphics[width=0.44\textwidth]{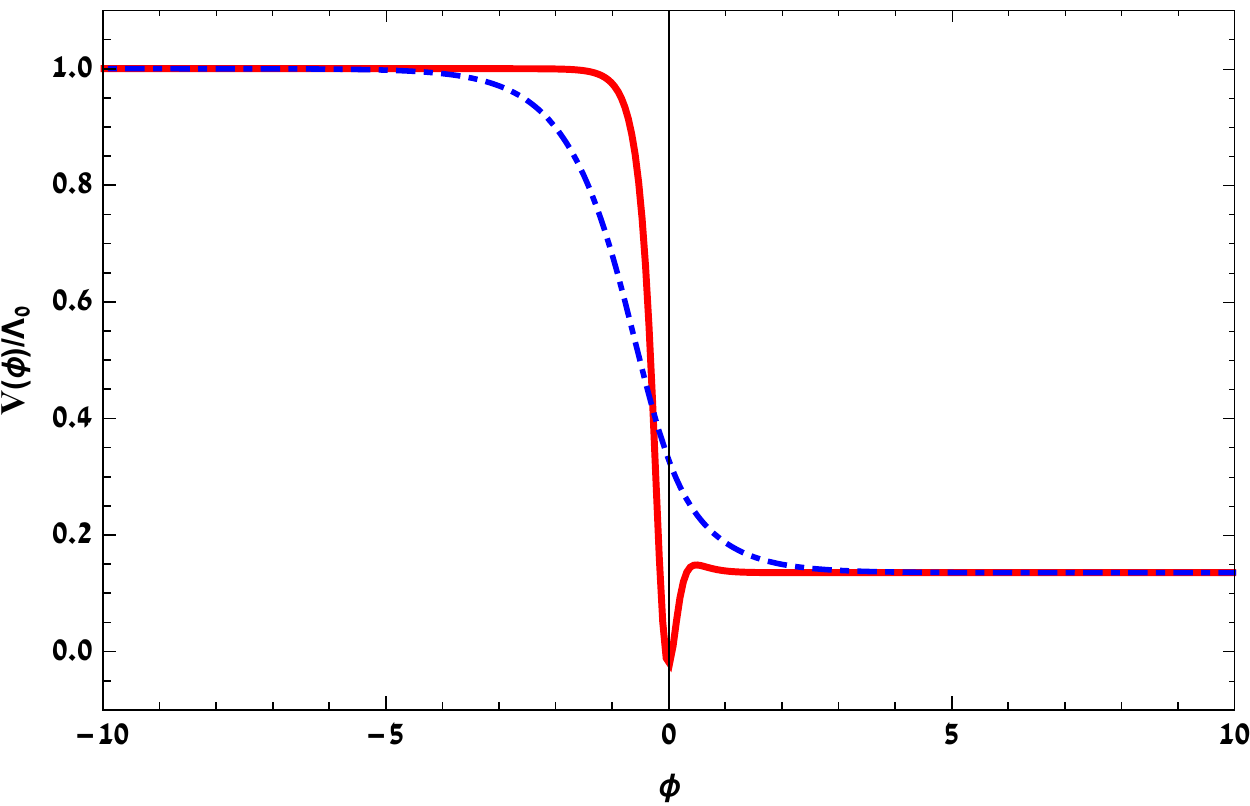}
\caption{{\it{The corresponding scalar field potential for the Lorentzian ansatz, with different values of $\Gamma$: $0.1$ red smooth line, $1$ blue dashed line.}}}
\label{fig3}
\end{figure}

Fig \ref{fig3} shows the scalar potential $V(\phi)$. The universe in this picture begins with $\phi \rightarrow \infty$ with a slow roll behavior and goes to the left-hand side. After approaching the minimum the universe evolves with another slow roll behavior that corresponds to the dark energy epoch when $\phi \rightarrow  - \infty$. The asymptotic values of the potential are:
\begin{equation}
V_{+\infty} = \Lambda_0 e^{\xi }, \quad V_{-\infty} = \Lambda_0 e^{-\xi } .
\end{equation}
Notice that this represents a see saw cosmological effect, that is if $\Lambda_0$ represents an intermediate scale, we see that in order to make the inflationary scale big forces the present vacuum energy to be small.  $\Lambda_0$ represents the geometric average of the inflationary vacuum energy and the present Dark Energy vacuum energies.
 For the limits $\Gamma \rightarrow 0$ the potential has a form of a step function for the two vacuum energies part. For the $\Gamma \sim 1$ the potential approaches the form of $tanh(x)$. Those features can be shown in Fig \ref{fig3}.

See saw cosmological effects in modified measure theories with spontaneously broken scale invariance have been studied in \cite{Guendelman:1999rj,Guendelman:1999qt,Guendelman:2014bva}. Notice that in the case studied here the smallness of the vacuum energy in the late universe is correlated with a large energy in the early universe, while in \cite{Guendelman:1999rj,Guendelman:1999qt,Guendelman:2014bva}. The small value of the vacuum energy in the late universe is related to a see saw formula ($\sim f_1^2/f_2$), where $f_1$  is of order of $M_{EW}^4$ and  $f_2$  is of order of $M_{Planck}^4$. In that case however both $f_1$ and $f_2$
are parameter defining a given flat region, so the see saw mechanisms are different.

 \section{Reheating}
\label{sec:reh}
The quintessential inflation scenarios does not allow for the standard oscillating part that creates particles \cite{Dimopoulos:2017tud,Agarwal:2017wxo,deHaro:2017nui,Lopez:2020zlu}. Therefore, in order to create particles we introduce a new scalar field $\sigma$ whose mass is time depended threw the field $\phi$. As an example:
\begin{equation}
m^2_\sigma = \mu^2 \exp\left[-2\frac{\alpha}{\Mpl} \phi \right]
\end{equation}
where $\mu$ has dimensions of mass and $\alpha$ determines the coupling of the inflaton with the $\sigma$ particles. Hence, the additional part of the action reads:
\begin{equation}
\mathcal{L}_\sigma = -\frac{1}{2} g^{\mu\nu} \sigma_{,\mu} \sigma_{,\nu} -  \frac{1}{2} m^2_\sigma \sigma^2.
\end{equation}
The creation of the $\sigma$ particles requires the time dependence of the inflaton field $\phi$ \cite{Haro:2018jtb}. The scalar field $\phi$ evolves according to the slow roll approximation up to the value $\phi_\text{end} = -6.41$. From this value of $\phi_\text{end}$ the inflaton falls into a state with a constant an  approximately constant velocity at the flat region $\dot{\phi}_0$ which can be calculated from the boundary conditions given at the end of inflation:
\begin{equation}
\dot{\phi}_0 =\sqrt{ \frac{\Gamma\xi }{\pi} }\frac{ H_0 e^{-\xi/2}}{N_{re}}
\end{equation}
where $N_{re}$ is the number of $e$ folding at the stage of reheating. The particle creation takes place when:
\begin{equation}
|\dot{m}_\sigma|\, > m_\sigma^2,
\end{equation}
that means it starts for 
\begin{equation}
|\dot{m}_\sigma|\, \approx m_\sigma^2.
\end{equation}
The condition yields:
\begin{equation}
\phi \approx - \frac{1}{\alpha} \log \left( \alpha \frac{|\dot{\phi}_0|}{\mu}\right).
\end{equation}
In order to estimate time creation of the $\sigma$ particles, we divide the typical field value where the particle creation starts  by the velocity:
\begin{equation}
\Delta t \sim - \phi/|\dot{\phi}_0| = \frac{1}{\alpha |\dot{\phi}_0| } \log \left( \alpha \frac{|\dot{\phi}_0|}{\mu}\right).
\end{equation}
From the time duration one can calculate the density of particles in the phase space \cite{Haro:2018jtb,Dey:2018mjg,Haro:2018zdb,Haro:2019gsv,deHaro:2019oki,Haro:2019ndy,Haro:2019peq,Haro:2019umj,Haro:2020rsj,Pan:2020mst}:
\begin{equation}
n_k  = \exp \left[- \pi k^2 \Delta t^2 \right],
\end{equation}
where we neglect the mass $m_\sigma$. $k$ is the momentum of the particle produce. The integration over the momentum gives the density:
\begin{equation}
n_\sigma = \frac{1}{2 \pi^2} \int_{0}^{\infty} dk k^2 n_k
= \frac{1}{8 \pi^3} \left[ \frac{\alpha |\dot{\phi}_0|} {\log (\alpha \dot{\phi}_0/\mu)}\right]^3 
\end{equation}
where we assume the mass $m_{\sigma}$ approaches zero, which is true for the specific dependence on the scalar field that we have chosen. This is the value for $a = a_0$. In order to generalize this quantity for an arbitrary scale parameters we use that the density deludes as $\sim a^3$:
\begin{equation}
n_\sigma = \frac{1}{8 \pi^3} \left(\frac{a_0}{a}\right)^3 \left[ \frac{\alpha |\dot{\phi}_0|} {\log (\alpha \dot{\phi}_0/\mu)}\right]^3.
\end{equation}
The energy density for the produced particles is:
\begin{equation}
\rho_\sigma = \frac{1}{\left(2\pi a\right)^3} \int_{0}^{\infty}n_k \sqrt{k^2/a^2 + m_\sigma^2} 4 \pi k^2 dk.
\end{equation}
For $\alpha$ big enough the $ \sqrt{k^2/a^2}$ decays slower than $m_\sigma^2$ and  the energy density of the sigma particles reads:
\begin{equation}
    \rho_\sigma = \left[\frac{\alpha \dot{\phi}_0}{\sqrt{2}\pi \log \left[ \alpha \dot{\phi_0} / \mu \right]}\right]^4 \frac{1}{a^4}
\end{equation}
that behaves as radiation. In order  to obtain a real thermal spectrum a thermalization process is required since the spectrum is not thermal. The thermalization should bring all particle species \cite{Mukaida:2015ria,Ahmad:2019jbm,Pinho:2020uzv}. 
This is just an example on how particles could be created after inflation. We now discuss some features that concern the evolution of the created mater that do not depend to much on the exact mechanism of particle creation. Similar calculations of particle creation for other quintessential inflationary models (obtained from spontaneous symmetry breaking of scale invariance) are being performed in \cite{Guendelman:2020zvt}, or from Supergravity Field redefinition \cite{Artymowski:2019jlh}.

\section{Late Time expansion}
\label{sec:late}
This section discuss the qualitative behavior for the matter fields, that include radiation and matter. The advantage of the this model in comparison to most quintessence models  is that the potential doesn't go to zero \cite{Zlatev:1998tr,Caldwell:1999ew,Chiba:1999ka,Bento:2002ps,Tsujikawa:2013fta}. Instead, the potential approaches a small constant values that represents the cosmological constant. This means that our model is very close to the $\Lambda$CDM from early times, as opposed to other quintessential potential where the potential approaches zero for large values. But still explains two epochs. For example deviations from a constant value of the effective potential can produce problems with structure formation \cite{Zimdahl:2019pqg}.

\begin{figure}[t!]
 	\centering
\includegraphics[width=0.44\textwidth]{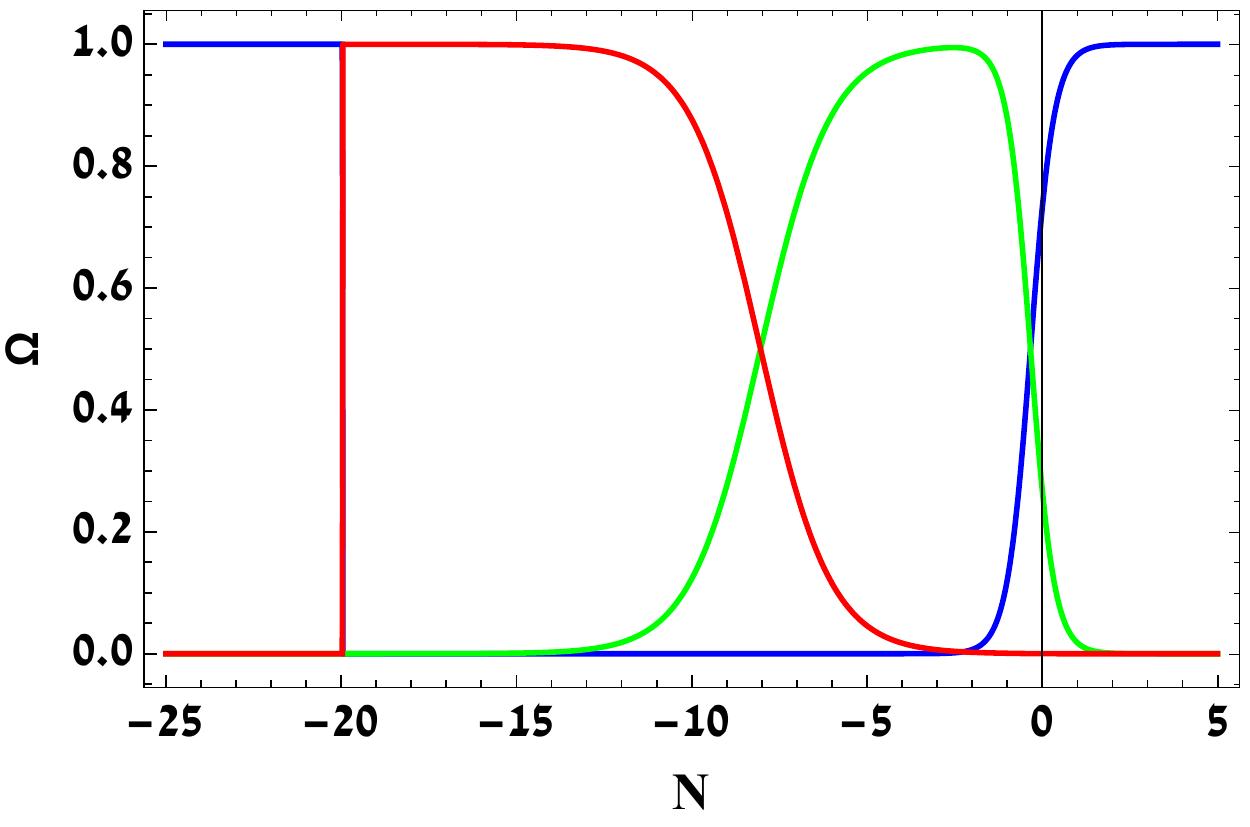}
\includegraphics[width=0.44\textwidth]{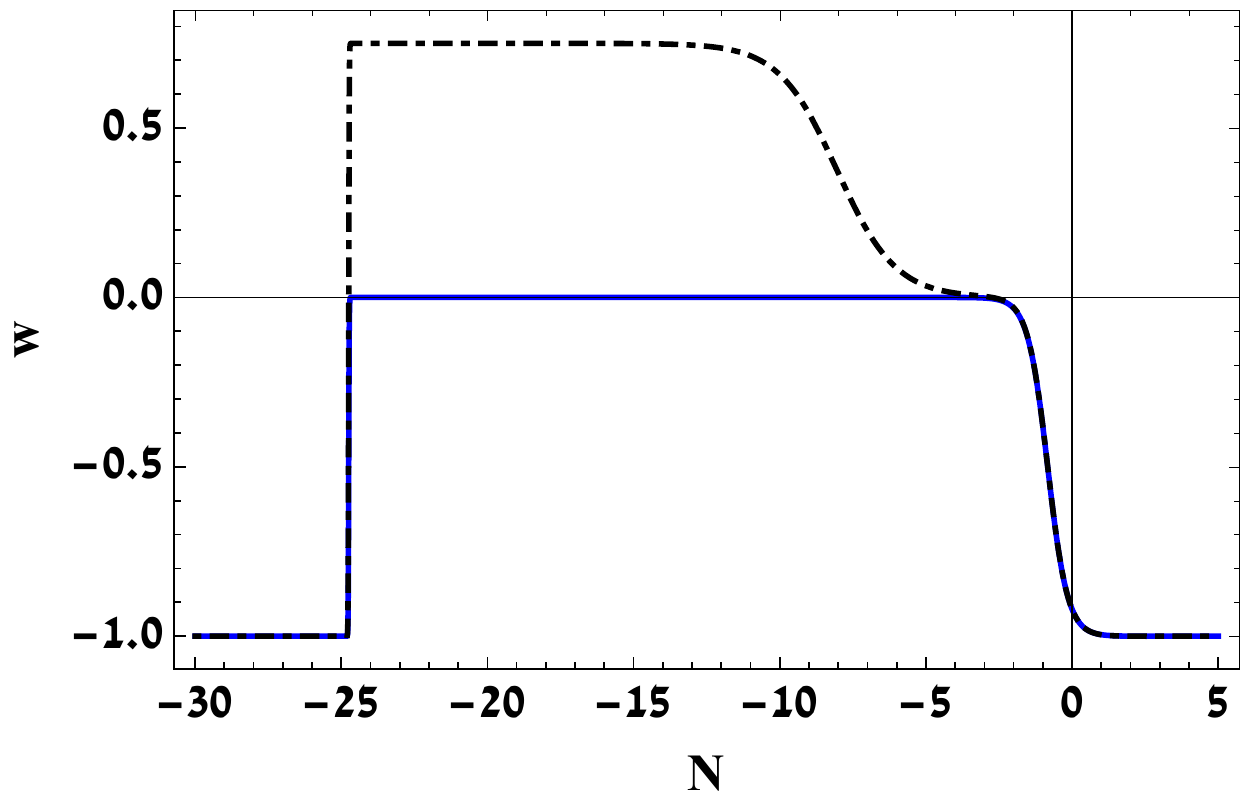}
\caption{{\it{The evolution of the energy densities for: $\Omega_m =  0.3089$,$\Omega_\Lambda = 0.691$,$\xi= 129$,$\Gamma = 0.1$. The center of the Lorentzian is around $N = -20$. Shortly after this point the radiation is created. The blue line is the energy density of the quintessential inflation. The red line is the energy density of the radiation, and the green line is for matter. In the lower panel: the blue line describes the evolution of the scalar field equation of states, and the blue dashed line is the total one.}}}
\label{fig4}
\end{figure}

We start with the action:
\begin{eqnarray}
&&\mathcal{S} = \int d^4x
\sqrt{-g}\bigg[\frac{\Mpl^2}{2}R-\frac{1}{2}\partial^\mu\phi\partial_\mu \phi-V(\phi)
\bigg]\nonumber\\
&&
+\S_\m+\S_\r ,
 \label{eq:action}
\end{eqnarray}
In flat Friedmann-Robertson-Walker (FRW) cosmology the two Friedmann equations read:
\begin{eqnarray}
3H^2\Mpl^2 = \frac{1}{2}\dot\phi^2+V(\phi)+\rho_\m+\rho_\r
 \label{eq:Fried1} \\
-\left(2\dot H+3H^2\right)\Mpl^2 = \frac{1}{2}\dot\phi^2-V(\phi)+\frac{1}{3}\rho_\r ,
 \label{eq:Fried2}
\end{eqnarray}
with:
\begin{equation}
 \ddot\phi+3H\dot\phi + V(\phi) = 0\, .
 \label{eq:eom_phi}
\end{equation}
and the matter fields read:
\begin{equation}
\rho_{m} = \frac{\Omega_m^{0}}{a^3}, \quad  \rho_{r} = \frac{\Omega_r^{0}}{a^4}.
 \label{eq:mat}
\end{equation}
Approximately, the exact solution in the absence of dust and radiation for the vacuum energy that we begin with, can be used directly in the Friedmann equations. In order to present the thermal history of the Universe in a more transparent way, we introduce the dimensionless density parameters for matter, radiation,
and scalar field, respectively given by
\begin{eqnarray}
 \Omega_m &=& \frac{\rho_m}{3H^2\Mpl^2} \, ,
 \label{eq:Omega_m}\\
   \Omega_r &=& \frac{\rho_r}{3H^2\Mpl^2} \, ,
 \label{eq:Omega_r}\\
  \Omega_\phi &=& \frac{\rho_\phi}{3H^2\Mpl^2} \, ,
 \label{eq:Omega_sig}
\end{eqnarray}
where $\rho_\phi=(1/2)\dot\phi^2+V(\phi)$, and in Fig. \ref{fig4} we depict their evolution as a function of the e-foldings. As we observe, we can reproduce the thermal history of the Universe, starting from a scalar field kinetic regime, then entering the radiation and matter regimes, and finally resulting to late-time dark-energy dominated era. Finally, for completeness, in Fig. \ref{fig4} we depict the corresponding behavior of the scalar field equation of state parameter $w_\phi\equiv p_\phi/\rho_\phi$, as well as of the effective equation of state parameter $w_{\rm eff}\equiv p_{\rm tot}/\rho_{\rm tot}=-1-2\dot{H}/3H^2$. From this figure we verify that around the present era the potential term dominates over the kinetic one in the scalar-field energy density, which leads $w_\phi$ to be around $-1$.

The Big Bang Nucleosynthesis (BBN) yields another constraint on the scalar inflaton model:
\begin{equation}
\Omega_\phi (z \sim 10^{10}) \preceq 0.09. 
\end{equation}
Considering the values: $\xi = 129,\Gamma = 0.1$ with the Planck values: $\Omega_m = 0.3089$, $\Omega_\Lambda = 0.691$ yields $\Omega_\phi \sim 10^{-36}$, which is deep in the BBN constraint.

\section{Conclusions}
\label{sec:con}
 
Here we formulate the slow roll parameter of the inflaton field as a Lorentzian form. From that assumption all of the higher order slow roll parameters are small and close to zero at the beginning of the inflationary period. Consequently, the inflationary observables agrees with the recent Planck observations. The ratio between the inflationary and the late time acceleration is determined by the value of $\xi \approx 129$, and the width of the transition is determined by $\Gamma \ll 10$.

In order to address in more detail scenario the matter fields contribution for the expansion of the universe, we example a reheating mechanism that is designed to work for the smooth potential obtained in our quintessential inflationary scenario that relies on the fast time dependence of some additional particles coupled to the dilaton field instead of oscillations of the inflation potential (which do not exist).

Independently on how the matter is obtained, the further evolution of the matter and vacuum energy is quite distinguishable from other quintessential inflationary scenarios, since the inflationary potential approaches a constant instead of approaching to zero. The models where the quintessential potential approaches zero has serious problems concerning structure formation see for example \cite{Zimdahl:2019pqg,Benisty:2020nql}.

In the future it would be interesting to investigate different localized forms for the first slow roll parameter. The Lorentzian form approaches a $\delta$ function as $\Gamma \rightarrow 0$. Therefore, different forms of $\epsilon$ that approach to delta function represent different transition forms from the inflationary and late time vacuum energy. For instance a Gaussian form for the $\epsilon $ parameter yields the same behavior.

\begin{acknowledgements}
This article is supported by COST Action CA 15117 "Cosmology and Astrophysics Network for Theoretical Advances and Training Action" (CANTATA) of the COST (European Cooperation in Science and Technology). This project is partially supported by COST Actions CA16104 and CA18108. D.B. thanks Ben-Gurion University of the Negev and Frankfurt Institute for Advanced Studies for generous support. D.B. thanks to Bulgarian National Science Fund for support via research grant KP-06-N 8/11.
\end{acknowledgements}

\bibliographystyle{spphys}       
\bibliography{ref}   
\end{document}